\documentclass[12pt]{article}
\usepackage{latexsym,amsmath}
\input amssym

\begin{document}
\renewcommand{\theequation}{\thesection.\arabic{equation}}
\def\prg#1{\medskip{\bf #1}}     \def\ra{\rightarrow}
\def\lra{\leftrightarrow}        \def\Ra{\Rightarrow}
\def\nin{\noindent}              \def\pd{\partial}
\def\dis{\displaystyle}          \def\dfrac{\dis\frac}
\def\grl{{GR$_\Lambda$}}         \def\vsm{\vspace{-9pt}}
\def\Lra{{\Leftrightarrow}}      \def\ads3{AdS$_3$}
\def\cs{{\scriptscriptstyle \rm CS}}  \def\ads3{{\rm AdS$_3$}}
\def\Leff{\hbox{$\mit\L_{\hspace{.6pt}\rm eff}\,$}}
\def\inn{\,\rfloor\,}
\def\bull{\raise.25ex\hbox{\vrule height.8ex width.8ex}}
\def\Tr{\hbox{\rm Tr\hspace{1pt}}}
\def\bF{{\bar F}}     \def\bt{{\bar\tau}}  \def\bg{{\bar g}}
\def\att#1{~{{\Large$\bullet$}~#1}}
\def\as{{\rm as}}     \def\inn{\,\rfloor\,}
\def\ric{{(Ric)}}     \def\tmgl{\hbox{TMG$_\Lambda$}}
\def\km{$u(1)_{\rm KM}$}
\def\kmv{$u(1)_{\scriptstyle KM}\oplus_{\scriptstyle sd}V$}

\def\bA{{\bar A}}     \def\bx{{\bar x}}     \def\bT{{\bar T}}
\def\bH{{\bar H}}     \def\bL{{\bar L}}     \def\bB{{\bar B}}
\def\hO{{\hat O}}     \def\hG{{\hat G}}     \def\tG{{\tilde G}}
\def\cL{{\cal L}}     \def\cM{{\cal M }}    \def\cE{{\cal E}}
\def\cA{{\cal A}}     \def\cI{{\cal I}}     \def\cC{{\cal C}}
\def\cF{{\cal F}}     \def\hcF{\hat{\cF}}   \def\bcF{{\bar\cF}}
\def\cH{{\cal H}}     \def\hcH{\hat{\cH}}   \def\bcH{{\bar\cH}}
\def\cK{{\cal K}}     \def\hcK{\hat{\cK}}   \def\bcK{{\bar\cK}}
\def\hcT{\hat{\cT}}   \def\bxi{{\bar\xi}}
\def\cO{{\cal O}}     \def\hcO{\hat{\cal O}} \def\tR{{\tilde R}}
\def\cB{{\cal B}}     \def\bV{{\bar V}}     \def\heps{\hat\epsilon}
\def\gp{{\goth p}}    \def\cT{{\cal T}}     \def\bcT{{\bar\cT}}
\def\hcO{{\hat\cO}}   \def\hcH{\hat\cH}     \def\bL{{\bar L}}
\def\bcH{\bar\cH}     \def\bcK{\bar\cK}     \def\bD{{\bar D}}
\def\bG{{\bar G}}     \def\bd{\bar\delta}   \def\D{{\Delta}}
\def\bu{{\bar u}}     \def\bv{{\bar v}}     \def\bw{{\bar w}}

\def\G{\Gamma}        \def\S{\Sigma}        \def\L{{\mit\Lambda}}
\def\a{\alpha}        \def\b{\beta}         \def\g{\gamma}
\def\d{\delta}        \def\m{\mu}           \def\n{\nu}
\def\th{\theta}       \def\k{\kappa}        \def\l{\lambda}
\def\vphi{\varphi}    \def\ve{\varepsilon}  \def\p{\pi}
\def\r{\rho}          \def\Om{\Omega}       \def\om{\omega}
\def\s{\sigma}        \def\t{\tau}          \def\eps{\epsilon}
\def\ups{\upsilon}    \def\tom{{\tilde\om}} \def\bw{{\bar w}}
\def\nab{\nabla}      \def\tnab{{\tilde\nabla}}
\def\tcR{\tilde{\cal R}} \def\tH{\tilde H}  \def\tG{\tilde G}
\def\Th{\Theta}       \def\cT{{\cal T}}     \def\cS{{\cal S}}
\def\nul{{\hat 0}}   \def\one{{\hat 1}}     \def\two{{\hat 2}}

\def\bm{{\bar m}}     \def\bn{{\bar n}}     \def\bk{{\bar k}}
\def\br{{\bar r}}     \def\bs{{\bar s}}
\def\nn{\nonumber}
\def\be{\begin{equation}}             \def\ee{\end{equation}}
\def\ba#1{\begin{array}{#1}}          \def\ea{\end{array}}
\def\bea{\begin{eqnarray} }           \def\eea{\end{eqnarray} }
\def\beann{\begin{eqnarray*} }        \def\eeann{\end{eqnarray*} }
\def\beal{\begin{eqalign}}            \def\eeal{\end{eqalign}}
\def\lab#1{\label{eq:#1}}             \def\eq#1{(\ref{eq:#1})}
\def\bsubeq{\begin{subequations}}     \def\esubeq{\end{subequations}}
\def\bitem{\begin{itemize}}           \def\eitem{\end{itemize}}

\title{Asymptotic structure of topologically massive gravity\\
       in spacelike stretched AdS sector}

\author{M. Blagojevi\'c and B. Cvetkovi\'c\footnote{
        Email addresses: {\tt mb@phy.bg.ac.yu,
                                cbranislav@phy.bg.ac.yu}} \\
University of Belgrade, Institute of Physics,\\ P. O. Box 57, 11001
Belgrade, Serbia}
\date{}
\maketitle
\begin{abstract}
We introduce a natural set of asymptotic conditions in the
spacelike stretched AdS sector of topologically massive gravity.
The Poisson bracket algebra of the canonical generators is shown
to have the form of the semi-direct sum of a $u(1)$ Kac-Moody and
a Virasoro algebra, with central charges. Using the Sugawara
construction, we prove that the asymptotic symmetry coincides with
the conformal symmetry, described by two independent Virasoro
algebras with central charges. The result is in complete agreement
with the hypothesis made in \cite{6}.

\end{abstract}

\section{Introduction}
\setcounter{equation}{0}

Topologically massive gravity with a cosmological constant $\L$,
denoted shortly as \tmgl, is an extension of three-dimensional general
relativity with a cosmological constant (\grl) by a gravitational
Chern-Simons term \cite{1}. While \grl\ is a topological theory, \tmgl\
is a dynamical theory with one propagating mode, the massive graviton.
In the AdS sector (with $\L<0$), \tmgl\ contains a maximally symmetric
vacuum solution, known as \ads3, and the related BTZ black hole
\cite{2}, with interesting thermodynamic properties. Thus, \tmgl\ seems
to be a useful model for exploring dynamical properties of the
gravitational dynamics. However, the interpretation of \tmgl\ for
generic values of the Chern-Simons coupling constant suffers from
serious difficulties: for the usual sign of the gravitational coupling
constant, $G>0$, massive excitations about \ads3\ carry negative
energy \cite{1}, while $G<0$ leads to the negative energy of the BTZ
black hole \cite{x1,3}.

In order to resolve this inconsistency, Li et. al. \cite{3}
introduced the so-called chiral version of the theory, defined by
a specific relation between the coupling constants, and argued
that it might lead to a consistent theory at both classical and
quantum level \cite{3,4}. Here, we follow another idea, related to
the fact that \tmgl\ has a rather rich vacuum structure
\cite{5,6}. Since the AdS sector of \tmgl\ around \ads3\ is not
consistent, Anninos et al \cite{6} proposed to choose a new
vacuum, the so-called spacelike stretched \ads3, which could be a
stable ground state of the theory \cite{7}. This choice reduces
the isometry group $SL(2,R)\times SL(2,R)$ of \ads3\  to its four
parameter subgroup $U(1)\times SL(2,R)$. Exploring thermodynamic
properties of the spacelike stretched black hole, Anninos et al
\cite{6} were led to a hypothesis that the corresponding boundary
dynamics is described by a {\it holographically dual
two-dimensional conformal field theory\/} (as in the standard
\ads3\ case). Recently, an extension of the above hypothesis to
the dS sector was discussed in \cite{8}.

As a natural step toward verification of the above hypothesis,
Comp\`ere et al \cite{9} (see also \cite{10}) investigated
asymptotic symmetries in the spacelike stretched \ads3\ sector.
They found a structure isomorphic to the semi-direct sum of the
$u(1)$ Kac-Moody algebra and the Virasoro algebra, \kmv, with a
central extension. The result looks quite natural, but the
validity of the hypothesis still remains an open issue.

In this paper, we also examine the correctness of the hypothesis
formulated in \cite{6}. Since we are convinced that the asymptotic
structure of a dynamical system is most clearly seen in the
canonical formalism, our approach is based on Dirac's constraint
Hamiltonian formalism, in a form applied recently to \tmgl\
\cite{11}. After formulating a set of natural asymptotic
conditions that generalize the usual AdS conditions \cite{12}, we
find that the asymptotic symmetry of the spacelike stretched AdS
sector of \tmgl\ is indeed a two-dimensional {\it conformal
symmetry\/} with central charges, in complete agreement with the
hypothesis of \cite{6}.

The paper is organized as follows. In section 2 we give a brief
overview of the basic dynamical features of \tmgl\ in the first-order
Lagrangian formalism and discuss the form of the spacelike stretched
black hole solution. Then, in section 3, we formulate the concept of
{\it asymptotically warped AdS configuration\/}, derive the asymptotic
conditions in the spacelike stretched sector, and find the form of the
corresponding asymptotic parameters. The asymptotic commutator algebra
is found to be \kmv\  {\it without\/} central charges. In section 4, we
study the canonical realization of the asymptotic symmetry by
constructing the Poisson bracket algebra of the improved canonical
generators. It turns out that it has the form \kmv\ {\it with\/}
central charges. This algebra is essentially of the same form as
the one found in \cite{9,10}. In section 5, we derive the main result
of this paper: using the Sugawara construction \cite{13} in the \km\
sector, we find that the asymptotic symmetry can be written in the form
of {\it two independent Virasoro algebras with central charges\/}, the
values of which coincide with those conjectured in \cite{6}. Section 6
is devoted to concluding remarks, while appendices contain some
technical details.

Our conventions are given by the following rules: the Latin indices
refer to the local Lorentz frame, the Greek indices refer to the
coordinate frame;  the middle alphabet letters
$(i,j,k,...;\m,\n,\l,...)$ run over 0,1,2, the first letters of the
Greek alphabet $(\a,\b,\g,...)$ run over 1,2; the metric components in
the local Lorentz frame are $\eta_{ij}=(+,-,-)$; totally antisymmetric
tensor $\ve^{ijk}$ and the related tensor density $\ve^{\m\n\r}$ are
both normalized as $\ve^{012}=1$.

\section{Spacelike stretched black holes} 
\setcounter{equation}{0}

Topologically massive gravity with a cosmological constant is
formulated as a gravitational theory in Riemannian spacetime. Instead
of using the standard Riemannian formalism, with an action defined in
terms of the metric, with find it more convenient to work in the
first-order formalism,  with the triad field and spin connection as
fundamental dynamical variables. Such an approach can be naturally
described in the framework of Poincar\'e gauge theory \cite{14}, where
basic gravitational variables are the triad field $b^i$ and the Lorentz
connection $A^{ij}=-A^{ji}$ (1-forms), and the corresponding field
strengths are the torsion $T^i$ and the curvature $R^{ij}$ (2-forms).
Using the notation $A^{ij}=:-\ve^{ij}{_k}\om^k$ and
$R^{ij}=:-\ve^{ij}{_k}R^k$, we have: $T^i=\nabla
b^i:=db^i+\ve^i{}_{jk}\om^j b^k$ and
$R^i=d\om^i+\frac{1}{2}\,\ve^i{}_{jk}\om^j\om^k$ (the wedge product
sign is omitted for simplicity).

The antisymmetry of $A^{ij}$ ensures that the underlying geometric
structure corresponds to Riemann-Cartan geometry, in which $b^i$ is
an orthonormal coframe, $g:=\eta_{ij}b^i\otimes b^j$ is the metric of
spacetime, $\om^i$ is the Cartan connection, and $T^i,R^i$ are the
torsion and the Cartan curvature, respectively. For $T_i=0$, this
geometry reduces to Riemannian.

\subsection{Lagrangian and the field equations}

The Lagrangian of \tmgl\ is defined by
\be
L=2ab^i R_i-\frac{\L}{3}\,\ve_{ijk}b^ib^jb^k\,
  +a\m^{-1}L_\cs(\om)+\l^i T_i\, ,                         \lab{2.1}
\ee
where $a=1/16\pi G$, $L_\cs(\om)=\om^i d\om_i
+\frac{1}{3}\ve_{ijk}\om^i\om^j\om^k$ is the Chern-Simons Lagrangian
for the Lorentz connection, and $\l^i$ (1-form) is the Lagrange
multiplier that ensures $T_i=0$. We assume that $G>0$, while the values
of $\m$ are generic. By construction, \tmgl\ is invariant under the local
Poincar\'e transformations:
\bea
\d_0 b^i{_\m}&=& -\ve^i{}_{jk}b^j{}_{\m}\th^k-(\pd_\m\xi^\r)b^i{_\r}
     -\xi^\r\pd_\r b^i{}_\m\, ,                            \nn\\
\d_0\om^i{_\m}&=& -\nabla_\m\th^i-(\pd_\m\xi^\r)\om^i{_\r}
     -\xi^\r\pd_\r\om^i{}_\m\, ,                           \nn\\
\d_0\l^i{_\m}&=& -\ve^i{}_{jk}\l^j{}_{\m}\th^k-(\pd_\m\xi^\r)\l^i{_\r}
     -\xi^\r\pd_\r\l^i{}_\m\, .                            \lab{2.2}
\eea

By varying the action $I=\int L$ with respect to $b^i,\om^i$ and
$\l^i$, one obtains the gravitational field equations. Using the third
equation $T_i=0$, which ensures that $\om^i$ is the Riemannian
(Levi-Civita) connection, the first two equations can be written as
\bsubeq\lab{2.3}
\bea
&&2aR_i-\L\ve_{ijk}b^jb^k+2a\m^{-1}C_i=0\, ,               \lab{2.3a}\\
&&\l_m=2a\m^{-1}L_{mn}b^n\, ,\qquad
  L_{mn}:=\ric_{mn}-\frac{1}{4}\eta_{mn}R\, .              \lab{2.3b}
\eea
\esubeq
Here, $C_i:=\nab (L_{ik}b^k)$ is the Cotton 2-form,
$\ric_{mn}=-\ve^{kl}{_m}R_{kln}$ and $R=-\ve^{ijk}R_{ijk}$. The
expansion in the basis $\heps_k=\frac{1}{2}\ve_{kmn}b^mb^n$, given by
$R_i=G^k{}_i\heps_k$, $C_i=C^k{}_i\heps_k$, yields the standard
component form of the first equation:
$$
aG_{ij}-\L\eta_{ij}+a\m^{-1}C_{ij}=0\, ,
$$
where $G_{ij}$ is the Einstein tensor, and $C_{ij}=\ve_i{}^{mn}\nab_m
L_{nj}$ the Cotton tensor.

\subsection{Construction of spacelike stretched black holes}

The spacelike stretched black hole \cite{x2,5,6} is a particular
solution of \tmgl\ with several attractive features: it is a
discrete quotient of the spacelike stretched \ads3\ vacuum
\eq{A1}, both solutions have the same type of asymptotic
behaviour, and the corresponding black hole thermodynamics
\cite{x2,15} seems to support the hypothesis made in \cite{6}, which
``predicts'' the existence of an asymptotic conformal symmetry in
this sector of \tmgl.

Using the results described in Appendix A, we are now going to
construct the spacelike stretched black hole in our first-order
formalism. After introducing a convenient notation,
$$
\L=-\frac{a}{\ell^2}\, ,\qquad \n=\frac{\m\ell}{3}\, ,
$$
we start from the spacelike stretched \ads3\ solution \eq{A1}, use
the coordinate transformations \eq{A3} and find the form of the
spacelike stretched black hole metric in Schwarzschild-like
coordinates $x^\m=(t,r,\vphi)$: \be
ds^2=N^2dt^2-B^{-2}dr^2-K^2(d\vphi+N_\vphi dt)^2\, , \lab{2.4} \ee
where \bea &&N^2=\frac{(\n^2+3)(r-r_+)(r-r_-)}{4K^2}\, ,\qquad
  B^2=\frac{4N^2K^2}{\ell^2}\, ,                           \nn\\
&&K^2=\frac{r}{4}\left[3(\n^2-1)r+(\n^2+3)(r_++r_-)
                 -4\n\sqrt{r_+r_-(\n^2+3)}\right]\, ,      \nn\\
&&N_\vphi=\frac{2\n r-\sqrt{r_+r_-(\n^2+3)}}{2K^2}\, .     \nn
\eea The metric of the spacelike stretched black hole \eq{2.4} is
defined for $\n^2>1$.

Going over to the first-order formalism, we choose the triad field
to have the simple diagonal form:
\bsubeq\lab{2.5}
\be
b^0=Ndt\, ,\qquad b^1=\frac{\ell}{2NK}dr\, ,\qquad
b^2=K(d\vphi+N_\vphi dt)\, .                               \lab{2.5a}
\ee
The connection $\om^i$ is determined by the condition of vanishing
torsion, $db^i+\ve_{ijk}\om^jb^k=0$:
\be
\om^0=-\b b^0-\g b^2\, ,\qquad \om^1=-\b b^1\, ,\qquad
\om^2=-\a b^0+\b b^2\, ,                                  \lab{2.5b}
\ee
where
$$
\a:=\frac{2KN'}{\ell}\, ,\qquad
\b:=\frac{K^2N_\vphi'}{\ell}\, ,\qquad \g:=\frac{2NK'}{\ell}\, .
$$
In the coordinate basis, we have:
\bea
&&\om^0=-\frac{N\n}\ell dt-\frac{2NKK'}{\ell}d\vphi\, ,\qquad
  \om^1=-\frac{KN_\vphi'}{2N}dr\, ,                        \nn\\
&&\om^2=-\frac{KN_\vphi\n}\ell dt
        +\frac{K^3N_\vphi'}\ell d\vphi\, .                 \nn
\eea
Finally, the solution for $\l_m$ takes the form:
\bea\lab{2.5c}
&&\l_0=2a\m^{-1}\left[\left((Ric)_{00}-\frac{3}{2\ell^2}\right)b^0
                      +(Ric)_{02}b^2\right]\, ,            \nn\\
&&\l_1=2a\m^{-1}\left((Ric)_{11}+\frac{3}{2\ell^2}\right)b^1\,,\nn\\
&&\l_2=2a\m^{-1}\left[(Ric)_{20}b^0
       +\left((Ric)_{22}+\frac{3}{2\ell^2}\right)b^2\right]\,,
\eea \esubeq where the Ricci tensor $\ric_{ij}$ is calculated in
Appendix B. Equations \eq{2.5} define the spacelike stretched
black hole in the first-order formalism.

\section{Asymptotic conditions} 
\setcounter{equation}{0}

In this section, we use a natural technique, known from earlier
studies of the AdS sector \cite{12}, to introduce asymptotic
conditions in the sector containing the spacelike stretched \ads3,
then we analyze the corresponding restrictions on the gauge
parameters and calculate the form of the commutator algebra.

\subsection{Spacelike stretched AdS asymptotics}

Let us introduce the concept of {\it warped AdS asymptotic
behavior}, based on the following requirements: \bitem \item[(a)]
asymptotic configurations should include warped black hole
geometries;\vsm \item[(b)] they should be invariant under the
action of $U(1)\times SL(2,R)$, the isometry group of warped
\ads3;\vsm \item[(c)] asymptotic symmetries should have
well-defined canonical generators. \eitem Here, we apply this
general concept to the case of spacelike stretched AdS
asymptotics.

The requirement (a) means that asymptotic conditions should be chosen
so as to include the spacelike stretched black hole configuration, defined
by \eq{2.5}.

In order to realize the requirement (b), we first consider the
spacelike stretched black hole metric \eq{2.4}. For large $r$, this metric
can be written in the form $g_{\m\n}=\bg_{\m\n}+\tG_{\m\n}$, where
$\bg_{\m\n}$ is the leading-order term (the black hole vacuum,
defined by $r_-=r_+=0$),
$$
\bar g_{\m\n}=
-\left(
\ba{ccc}
  1   &0      &\n r\\
  0   &\dis\frac{\ell^2}{(\n^2+3)r^2} &0\\
  \n r&0  &\dis\frac{3}{4}(\n^2-1)r^2
\ea
\right)\, ,
$$
and $\tG_{\m\n}$ represents the sub-leading terms. Let us now act on
$g_{\m\n}$ with all possible isometries of the spacelike warped \ads3,
defined by the four Killing vectors
$\xi_K=(\xi_{(2)},\bxi_{(1)},\bxi_{(2)},\bxi_{(0)})$, displayed in
appendix A. The result of this procedure has the form
$$
\d_K g_{\m\n}=\left(
\ba{ccc}
  0&\cO_2&\cO_0\\
  \cO_2&\cO_3&\cO_1\\
  \cO_0&\cO_1&\cO_{-1}
\ea
\right)\, ,
$$
where $\cO_n$ is a quantity that tends to zero as $1/r^n$ or faster
when $r\ra\infty$. In order to have a set of asymptotic configurations
which is sufficiently large to include the whole family of the metric
configurations $\bg_{\m\n}+\d_K g_{\m\n}$, as required by (b), we
adopt the following asymptotic form of the metric:
\be
g_{\m\n}=\bg_{\m\n}+G_{\m\n}\, , \qquad
G_{\m\n}= \left(
\ba{ccc}
  \cO_1& \cO_2& \cO_0 \\
  \cO_2& \cO_3& \cO_1 \\
  \cO_0& \cO_1& \cO_{-1}
\ea
\right)\, .                                                \lab{3.1}
\ee
Comparing this result with  \cite{10}, we find a complete agreement.

To simplify further discussion, we will use the notation
$\bar\phi:=\phi(r_-=r_+=0)$ for the leading-order term of any dynamical
variable $\phi$, which is a natural extension of the notation used for
the metric.

Although  metric is not a dynamical variable in our first order
formalism, its asymptotic form can be used to ``derive'' asymptotic
behaviour of the triad field. Indeed, by combining \eq{2.5a} and
\eq{3.1}, we are led to adopt the following asymptotic form of the triad
field:
\bsubeq\lab{3.2}
\be
b^i{_\m}=\bar b^i{_\m}+B^i{_\m}\, ,\qquad
B^i{_\m}:=\left(
          \ba{ccc}
          \cO_1&\cO_2&\cO_2\\
          \cO_2&\cO_2&\cO_1\\
          \cO_1&\cO_2&\cO_0
          \ea
          \right)\, .                                      \lab{3.2a}
\ee
Similarly, we combine \eq{3.2a} with \eq{2.5b} to find
\be
\om^i{_\m}=\bar\om^i{_\m}+\Om^i{_\m}\, ,\qquad
\Om^i{_\m}:=\left(
            \ba{ccc}
            \cO_1&\cO_2&\cO_0\\
            \cO_2&\cO_2&\cO_1\\
            \cO_1&\cO_2&\cO_0
            \ea
            \right)\, .                                    \lab{3.2b}
\ee
Finally, by combining \eq{3.2a} with \eq{2.5c}, we obtain:
\be
\l^i{_\m}= \bar\l^i{_\m}+\L^i{_\m}\, ,\qquad
\L^i{_\m}:=\left(
           \ba{ccc}
           \cO_1&\cO_2&\cO_0\\
           \cO_2&\cO_2&\cO_1\\
           \cO_1&\cO_2&\cO_0
           \ea
           \right)\, .                                     \lab{3.2c}
\ee
\esubeq

One should note that the asymptotic conditions are not uniquely
determined by the requirements (a) and (b). In the above procedure, we
were looking for the most general asymptotic behaviour compatible with
(a) and (b), with arbitrary higher-order terms. Later, when we consider
the condition (c), certain relations among the higher-order terms will
be established (in Appendix C).

By construction, the adopted asymptotic conditions are invariant under
the action of the isometry group $U(1)\times SL(2,R)$ of the spacelike
warped \ads3. Now, we wish to clarify the symmetry structure of the
field configurations \eq{3.2}.

\subsection{Asymptotic parameters}

Having chosen the asymptotic conditions \eq{3.2}, we are now going to
find the subset of gauge transformations \eq{2.2} that leave these
conditions invariant. More precisely, acting on the fields \eq{3.2},
these restricted (or asymptotic) gauge transformations are, by
definition, allowed to change only the (arbitrary) higher-order terms.
Consequently, the restricted gauge parameters are defined by the relations
\bea
&&-\ve^i{}_{jk}b^j{}_{\m}\th^k-(\pd_\m\xi^\r)b^i{_\r}
                -\xi^\r\pd_\r b^i{}_\m=\d_0 B^i{_\m}\, ,   \nn\\
&&-(\pd_\m\th^i+\ve^i{}_{jk}\om^j{_\m}\th^k)
  -(\pd_\m\xi^\r)\om^i{_\r}-\xi^\r\pd_\r\om^i{}_\m
                                      =\d_0\Om^i{_\m}\, ,  \nn\\
&&-\ve^i{}_{jk}\l^j{}_{\m}\th^k-(\pd_\m\xi^\r)\l^i{_\r}
                -\xi^\r\pd_\r\l^i{}_\m=\d_0\L^i{_\m}\, .   \nn
\eea
By solving these equations, we find the asymptotic parameters
for local translations,
\bsubeq\lab{3.3}
\bea
&&\xi^0=\ell T(\vphi)+\cO_2\,,\qquad \xi^1=-r\pd_2 S(\vphi)
        +\cO_0\, ,                                         \nn\\
&&\xi^2=S(\vphi)+\cO_2\, ,
\eea
and for local Lorentz rotations:
\bea
&&\th^0=-\frac{2\ell}{\sqrt{3(\n^2+3)(\n^2-1)}r}\pd_2^2S(\vphi)
        +\cO_2\, ,                                         \nn\\
&&\th^1=\frac{2\ell \sqrt{\n^2+3}}{3(\n^2-1)r}\pd_2 T(\vphi)
        +\cO_3\, ,                                         \nn\\
&&\th^2=-\frac{4\ell\n}{(\n^2+3)\sqrt{3(\n^2-1)}}
         \frac{1}{r}\pd_2^2S(\vphi)+\cO_2\, .
\eea
\esubeq
These parameters define the symmetry of the (asymptotic) boundary of
spacetime, in the spacelike stretched AdS sector of \tmgl.

\subsection{Asymptotic symmetry}

To find the interpretation of the asymptotic parameters, we calculate
the commutator algebra of the corresponding gauge transformations. To
begin with, we observe that commutator algebra of the local Poincar\'e
transformations \eq{2.2} is closed: $[\d_0(1),\d_0(2)]=\d_0[3]$, where
$\d_0(1):=\d_0(\xi^m_1,\th^i_1)$ etc, while the composition rule is
given by:
\bea
&&\xi^\m_3=\xi_1\cdot\pd\xi^\m_2-\xi_2\cdot\pd\xi^\m_1\, , \nn\\
&&\th^i_3=\ve^i{}_{mn}\th^m_1\th^n_2+\xi_1\cdot\pd\th_2^i
          -\xi_2\cdot\pd\th_1^i\, .                        \nn
\eea
Substituting here the asymptotic parameters \eq{3.3} and comparing the
lowest order terms, we obtain:
\bea
&&T_3=S_1\pd_2T_2-S_2\pd_2T_1\, ,                         \nn\\
&&S_3=S_1\pd_2S_2-S_2\pd_2S_1\, .                         \lab{3.4}
\eea
To clarify the meaning of this result, it is useful to define the
residual or pure gauge transformations as the transformations generated
by the higher order terms in \eq{3.3}. Pure gauge transformations are
known to be irrelevant in the canonical analysis of the asymptotic
structure of spacetime \cite{16}. This fact is made more precise by
saying that the {\it asymptotic symmetry group\/} is defined as the
factor group of gauge transformations generated by \eq{3.3}, with
respect to the residual gauge transformations. In other words, the
asymptotic symmetry is defined by the pair $(T,S)$, ignoring all the
residual, higher-order terms.

Now, introducing the Fourier expansion of the parameters and the
related notation
\bea
&&k_n:=\d_0(T=e^{in\vphi},S=0)\, ,                         \nn\\
&&\ell_n:=\d_0(T=0,S=e^{in\vphi})\, ,                      \nn
\eea
the commutator algebra of the asymptotic transformations takes
the form the semi-direct sum of \km\ and the Virasoro algebra,
\bea
&&i\left[k_m,k_n\right]=0\, ,                              \nn\\
&&i\left[k_m,\ell_n\right]=mk_{m+n}\, ,                    \nn\\
&&i\left[\ell_m,\ell_n\right]=(m-n)\ell_{m+n}\, .          \lab{3.5}
\eea
The same algebra was also found in \cite{9,10}. Central
charges are here absent, but they will appear in the canonical
analysis.

The adopted asymptotic conditions \eq{3.2} are chosen in agreement with
the requirements (a) and (b), formulated at the beginning of this
section, and the related symmetry structure is encoded in the form of
the asymptotic gauge parameters \eq{3.3}. The status of the requirement
(c) will be clarified in the canonical analysis of the next section.

\section{Canonical realization of the asymptotic symmetry} 
\setcounter{equation}{0}

Asymptotic symmetry of a gauge theory is most clearly understood in
the framework of the canonical formalism.  In this section, we apply
the results obtained in \cite{11} to study the canonical aspects of the
asymptotic structure of \tmgl\ in the spacelike stretched AdS sector.

Using the Castellani algorithm \cite{17}, we found the following
expression for the canonical gauge generator \cite{11}:
\bea
G&=&-G_1-G_2\,,\nn\\
G_1&=&\dot\xi^\r\left(b^i{}_\r\pi_i{}^0
       +\l^i{_\r}p_i{}^0+\om^i{}_\r\Pi_i{}^0\right)        \nn\\
&&+\xi^\r\left[b^i{}_\r\bcH_i+\l^i{_\r}\bcT_i
  +\om^i{}_\r\bcK_i+(\pd_\r b^i_0)\pi_i{}^0
  +(\pd_\r\l^i{_0})p_i{^0}+(\pd_\r\om^i{}_0)\Pi_i{^0}\right]\, ,\nn\\
G_2&=&\dot{\th^i}\Pi_i{}^0+\th^i\left[\bcK_i
  -\ve_{ijk}\left(b^j{}_0\pi^{k0}+\l^j{}_0p^{k0}
  +\om^j{}_0\Pi^{k0}\right)\right]\, .                     \lab{4.1}
\eea
Here, the integration symbol $\int d^3x$ is omitted for simplicity,
the canonical momenta corresponding to $(b^i{_\m},\om^i{_\m},\l^i{_\m})$,
are denoted as $(\pi_i{^\m},\Pi_i{^\m},p_i{^\m})$, and explicit expressions
for various terms appearing in $G$ are given in Appendix D.
The action of the gauge generator $G$ on the fields, defined
by $\d_0\phi=\{\phi,G\}$, has the form \eq{2.2}.

\subsection{Surface terms}

Since canonical generators act on dynamical variables via the Poisson
bracket (PB) operation, they must have well-defined functional
derivatives. When this is not the case, the problem can be usually
solved by adding suitable surface terms \cite{18}.

We start by examining the variation of the Lorentz generator
$G_2$:
\bea
\d G_2&=&\th^i\d \cK_i+\pd\hcO+R                           \nn\\
&=&-2a\ve^{0\a\b}\pd_\a(\th^i \d b_{i\b}
   +\th^i\m^{-1}\d\om_{i\b})+\pd\hcO+R                     \nn\\
&=&\pd\cO_1+R\, .                                          \nn
\eea
Here, $\hat\cO$ are terms with arbitrarily fast asymptotic decrease,
$R$ are regular terms which do not contain variations of the
derivatives of fields, and the final result is a consequence of the
asymptotic conditions \eq{3.2}. Since both $\cO_1$ and $R$ terms do not
contribute to surface integrals, it follows that $G_2$ is a
well-defined generator.

For $G_1$, we have:
\bea
\d G_1&=&\xi^\r\left(b^i{_\r}\d\cH_i+\om^i{_\r}\d \cK_i
                     +\l^i{_\r}\d\cT_i\right)+\pd\hcO+R    \nn\\
 &=&-\ve^{0\a\b}\pd_\a\Bigl[\xi^\r b^i{_\r}
    \left(2a\d\om_{i\b}+\d\l_{i\b}\right)                  \nn\\
&& +\xi^\r\l^i{_\r}\d b_{i\b}
   +\xi^\r\om^i{_\r}(2a\d b_{i\b}
   +2a\m^{-1}\d\om_{i\b})\Bigr]+\pd\hcO+R\, .              \nn
\eea
Using the adopted asymptotic conditions, we find:
\bea
\d G_1&=&- \pd_\a(\xi^0\d\cE^\a+\xi^2\d\cM^\a)+\cO_1+R     \nn\\
&=&-\d\pd_\a(\xi^0\cE^\a+\xi^2\cM^\a)+\cO_1+R\, ,          \nn
\eea
where
\bea
&&\cE^\a=\ve^{0\a\b}\left[b^0{_0}\left(\frac{4a}{3}\om^{0}{_\b}
  +\l^0{_\b}\right)-b^2{_0}\left(\frac{4a}{3}\om^2{_\b}+\l^2{_\b}
  -\frac{a}{3\ell}\frac{(2\n^2+3)}{\n}b^2{_\b}\right)\right]\, ,\nn\\
&&\cM^\a=-\ve^{0\a\b}\left[b^2{_2}(2a\om^2{_\b}+\l^2{_\b})
  +\frac{a\ell}{3\n}(\om^2{_2}\om^2{_\b}-\om^0{_2}\om^0{_\b})\right]\, .
\eea
Then, after re-introducing the spatial integration, the improved
generator takes the form
\bea
&&\tG=G+\G\, ,                                             \nn\\
&&\G:=-\int_0^{2\pi}d\vphi\left(\ell T\cE^1+S\cM^1\right)\,.\lab{4.3}
\eea

\subsection{Energy and angular momentum}

The general relation \eq{4.3} implies:
$$
\tG[\xi^0]=G[\xi^0]-\G[\xi^0]\, ,\qquad
\tG[\xi^2]=G[\xi^2]-\G[\xi^2]\, .
$$
For $\xi^0=1$ and $\xi^2=1$, the values of the surface terms have
the meaning of energy and angular momentum of the system,
respectively:
\be
E=\int_0^{2\pi}d\vphi\,\cE^1 \, ,\qquad
M=\int_0^{2\pi}d\vphi\,\cM^1 \, .                          \lab{4.5}
\ee

Let us show that these expressions are {\it finite\/}. Using the
adopted asymptotic conditions, one can express $\cE^1$ and $\cM^1$ as
functions of the sub-leading terms $(B^i{_\m},\Om^i{_\m},\L^i{_\m})$:
\bea
\cE^1&=&\frac{4a}{3}\sqrt{\frac{\n^2+3}{3(\n^2-1)}}
  \left(\Om^0{_2}+\frac{3}{4a}\L^0{_2}\right)              \nn\\
&&+\frac{8a\n}{3\sqrt{3(\n^2-1)}}\left(
  \frac{2\n^2+3}{4\n}\frac{B^2{_2}}\ell
  -\Om^2{_2}-\frac{3}{4a}\L^2{_2}\right)+\cO_1\, ,         \nn\\
\cM^1&=&-a\left[B^2{_2}\left(2\Om^2{_2}+\frac{1}{a}\L^2{_2}\right)
  +\frac{\ell}{3\n}(\Om^2{_2}-\Om^0{_2})(\Om^2{_2}+\Om^0{_2})\right]\nn\\
&&-\frac{a\sqrt{3(\n^2-1)}}{2}\left[\frac{B^2_2}{\n\ell}
  +\frac{4}{3}\Om^2{_2}+\frac{1}{a}\L^2{_2}
  +\frac{2\sqrt{\n^2+3}}{3\n}\Om^0{_2}\right]r+\cO_1\, .   \nn
\eea
Since the sub-leading terms are either constant or tend to zero in the
asymptotic region, it follows immediately that $\cE^1=\cO_0$, and
consequently, the expression for $E$ in \eq{4.5} is finite. In order to
prove the finiteness of the angular momentum, we need the improved
asymptotic relation \eq{C3}, derived in Appendix C. It implies
$\cM^1=\cO_0$, which completes the proof of finiteness.

Now, we can calculate energy and angular momentum of the spacelike
stretched black hole \eq{2.4}:
\bea
E&=&\frac{(\n^2+3)}{24G\ell}\left[r_++r_-
    -\frac{1}{\n}\sqrt{r_+r_-(3+\n^2)}\right]\,            \nn\\
M&=&-\frac{\n^2+3}{384G\ell\n}\left[(\n^2+3)(r_++r_-)
  +8(r_++r_-)\n\sqrt{r_+r_-(3+\n^2)}-2r_+r_-(11\n^2+9)\right]\nn\\
&\equiv&\frac{\n(\n^2+3)}{96G\ell}\left[\left(r_++r_-
  -\frac{1}{\n}\sqrt{r_+r_-(3+\n^2)}\right)^2
  -\frac{5\n^2+3}{4\n^2}(r_+-r_-)^2\right]\, ,
\eea
The result coincides with the ADT charges that can be found in \cite{6}
(see also \cite{x2,19}).

Returning now to the beginning of this section, where we introduced the
concept of the warped AdS asymptotics, we see that our asymptotic
conditions \eq{3.2} are also in agreement with the last requirement (c).

\subsection{Canonical algebra}

Now, we wish to find the PB algebra of the improved canonical
generators.

After introducing the notation $\tG(1):=\tG[T_1,S_1]$,
$\tG(2):=\tG[T_2,S_2]$, we use the main theorem of \cite{20} to
conclude that the PB $\{\tG(2),\tG(1)\}$ of two differentiable
generators is also a differentiable generator. This implies
\bsubeq\lab{4.7}
\be
\left\{\tG(2),\tG(1)\right\}= \tG(3)+C_{(3)},             \lab{4.7a}
\ee
where the parameters of $\tG(3)$ are defined by the composition rule
\eq{3.4}, while $C_{(3)}$ is an unknown field-independent functional,
$C_{(3)}:= C_{(3)}[T_1,S_1;T_2,S_2]$, the central term of the canonical
algebra. The form of $C_{(3)}$ can be found using the relation
$$
\d_0(1)\G(2)\approx \G(3)+C_{(3)}\, ,
$$
which is a consequence of $\{\tG(2),\tG(1)\}\approx\d_0(1)\G(2)$. The
expression $\d_0(1)\G(2)$ is calculated using the transformation laws
\bea
&&\d_0\cE^1=-S\pd_2\cE^1-(\pd_2 S)\cE^1
  -\frac{2a(\n^2+3)}{3\n}\pd_2 T\, ,                       \nn\\
&&\d_0\cM^1=-2(\pd_2 S)\cM^1-S\pd_2\cM^1-(\ell\pd_2 T)\cE^1
  -\frac{2a\ell(5\n^2+3)}{3\n(\n^2+3)}\pd_2^3S\, .         \nn
\eea
Once we know $\d_0(1)\G(2)$, we can identify the central term:
\be
C_{(3)}=\frac{2a\ell(\n^2+3)}{3\n}\int_0^{2\pi}d\vphi T_2\pd_2 T_1
  +\frac{2a\ell(5\n^2+3)}{3\n(\n^2+3)}
                      \int_0^{2\pi}d\vphi S_2\pd_2^3S_1\, .\lab{4.7b}
\ee
\esubeq

The form of the canonical algebra \eq{4.7} implies that the improved
generator is {\it conserved}. Indeed, using the relation
$\tG[1,0]=-\ell\tH_T$ and the composition rule \eq{3.4}, we have:
\bea
\frac{d}{dt}\tG&=&\frac{\pd}{\pd t}\tG+\{\tG,\tH_T\}     \nn\\
&=&\frac{\pd}{\pd t}\tG-\frac{1}{\ell}\{\tG[T,S],\tG[1,0]\}
  \approx \frac{\pd}{\pd t}\G[T,S]=0\, ,                 \nn
\eea
since the parameters $T$ and $S$ are time independent. Consequently, we
have the conservation of the surface term $\G$, and hence, the
conservation of the energy and the angular momentum.

After expressing the canonical generator in terms of the Fourier modes,
$$
K_n:=\tG(T=e^{-in\vphi},S=0)\, ,\qquad
L_n:=\tG(T=0,S=e^{-in\vphi})\, ,
$$
the canonical algebra \eq{4.7} takes a more familiar form:
\bsubeq\lab{4.8}
\bea
&&i\{K_m,K_n\}=-\frac{c_K}{12}m\d_{m,-n}\, ,                \nn\\
&&i\{K_m,L_n\}=mK_{m+n}\, ,                                \nn\\
&&i\{L_m,L_n\}=(m-n)L_{m+n}+\frac{c_V}{12}m^3\d_{m,-n}\, , \lab{4.8a}
\eea
where
\be
c_K=\frac{(\n^2+3)\ell}{G\n}\, ,\qquad
c_V=\frac{(5\n^2+3)\ell}{G\n(\n^2+3)}\, .                  \lab{4.8b}
\ee
\esubeq
Thus, the canonical realization of the asymptotic symmetry is given as
the semi-direct sum of $u(1)_{\rm KM}$ and the Virasoro algebra, with
central charges $c_K$ and $c_V$.

The authors of \cite{9,10} used different methods in their study of the
boundary conditions of the warped AdS sector of \tmgl. By comparing our
canonical result \eq{4.8} with equation (25) in \cite{10}, one finds
essentially a complete agreement, up to some differences in
conventions, stemming from different normalizations of $K_m$'s and a
shift in the value of $K_0$.

\section{Sugawara construction} 
\setcounter{equation}{0}

Clearly, the asymptotic algebra \eq{4.8} does not describe the
conformal symmetry, as conjectured in \cite{6}. However,
there is a particular construction due to Sugawara \cite{13}, which
reveals how the conformal algebra can be reconstructed on the basis
of \eq{4.8}. In this procedure, the presence of central charges
is of essential importance.

In the first step, we introduce the set of generators
\bsubeq
\be
\bL_n:=-\frac{6}{c_K}\sum_r K_rK_{n-r}\, ,
\ee
which obey the following PB relations:
\bea
&&i\{K_m,\bL_n\}=mK_{m+n}\,,                               \nn\\
&&i\{\bL_m,\bL_n\}=(m-n)\bL_{m+n}\,,                       \nn\\
&&i\{\bL_m,L_n\}=(m-n)\bL_{m+n}\, .                        \nn
\eea
Next, we introduce
\be
L_n^-:=L_n-\bL_n\, ,
\ee
whereupon \eq{4.8} takes the form of a \textit{direct sum} of
\km\ and the Virasoro algebra:
\bea
&&i\{K_m,K_n\}=-\frac{c_K}{12}m\d_{m,-n}\,,                \nn\\
&&i\{K_m,L^-_n\}=0\, ,                                     \nn\\
&&i\{L_m^-,L_n^-\}=(m-n)L^-_{m+n}
  +\frac{c^-}{12}m^3\d_{m,-n}\,,                           \nn
\eea
where $c^-:=c_V$. Finally, we define
\be
L_n^+:=-\bL_{-n}-in\a K_{-n}\, ,
\ee
\esubeq
where $\a$ is an arbitrary constant. The PB algebra between
$L^\mp_n$ takes the well-known form:
\bea
&&i\{L_m^+,L_n^+\}=(m-n)L_{m+n}^++\frac{c^+}{12}m^3\d_{m,-n}\,,\nn\\
&&i\{L_m^+,L_n^-\}=0\, ,                                   \nn\\
&&i\{L_m^-,L_n^-\}=(m-n)L^-_{m+n}+\frac{c^-}{12}m^3\d_{m,-n}\,,\lab{5.2}
\eea
where $c^+:=c_K\a^2$. This result reveals the conformal structure
hidden in \eq{4.8}.

Clearly, the value of $\a$ in $c^+$ has to be fixed by some additional
requirements. Before going to that, we display here the values of
$L_0^\pm$ in terms of the canonical energy and angular momentum for
the spacelike warped black hole:
\bea
&&L_0^+=\frac{6G\n\ell}{\n^2+3} E^2=\frac{(\n^2+3)\n}{96G\ell}
  \left[r_++r_--\frac 1\n\sqrt{r_+r_-(3+\n^2)}\right]^2\,,\nn\\
&&L_0^-=L_0^+-M=\frac{(\n^2+3)(5\n^2+3)}{384\n G\ell}(r_+-r_-)^2\, .
\eea
These results are in complete agreement with those found in \cite{6}.

In order to find out the value of $\a$, one can use our central charges
$c^\mp$ to calculate the black hole entropy via Cardy's formula:
$$
S_{\rm c}=2\pi\sqrt{\frac{L_0^+c^+}{6}}+2\pi\sqrt{\frac{L_0^-c^-}{6}}\, .
$$
A direct calculation leads to
\be
S_{\rm c}=2\pi\a\frac{(\n^2+3)}{24G}
  \left[r_++r_--\frac1\n\sqrt{r_+r_-(3+\n^2)}\right]
  +\frac{\pi(5\n^2+3)}{24\n G}(r_+-r_-)\, .
\ee
On the other hand, the gravitational black hole entropy of \tmgl\ has
the form \cite{15,6}:
\be
S_{\rm gr}=\frac{\pi}{24\n G}\left[(9\n^2+3)r_+-(\n^2+3)r_-
           -4\n\sqrt{r_+r_-(\n^2+3)}\right]\, .            \lab{5.5}
\ee
Comparing $S_{\rm c}$ and $S_{\rm gr}$, one finds that $S_{\rm
c}=S_{\rm gr}$ for
\be
\a=\frac{2\n}{\n^2+3}\equiv \frac{2\ell}{Gc_K}\, .         \lab{6.7a}
\ee
Consequently, the values of the central charges in the Virasoro
algebras \eq{5.2} are the same as those conjectured in \cite{6}:
\be
c^-=\frac{(5\n^2+3)\ell}{G\n(\n^2+3)}\, ,\qquad
c^+=\frac{4\n\ell}{G(\n^2+3)}\, .                          \lab{5.7}
\ee

In conclusion, our main result is expressed by the formulas \eq{5.2}
and \eq{5.7}, and it confirms the hypothesis formulated heuristically
in \cite{6}, at least at the classical level.

\section{Concluding remarks} 

In this paper, we analyzed asymptotic structure of \tmgl\ in the
spacelike stretched AdS sector.

(1) We introduced spacelike stretched AdS asymptotic conditions and
found the form of the corresponding asymptotic parameters. The
commutator algebra of the asymptotic transformations is the semi-direct
sum of \km\ with the Virasoro algebra, without central charges, which
is a natural generalization of the vacuum isometry algebra $u(1)\oplus
sl(2,R)$. Asymptotic conditions for the metric recently proposed in
\cite{10} coincide with ours.

(2) With the adopted asymptotic conditions, we constructed the improved
canonical generators and found the expressions for the conserved
charges. In particular, we calculated the energy and angular momentum
of the spacelike stretched black hole. We showed that canonical algebra
of the improved generators takes the form of the semi-direct product of
\km\ and the Virasoro algebra, with two central charges. Our algebra
has essentially the same form as the one found in \cite{9,10} by
different methods.

(3) In the last step, we used the Sugawara construction in the \km\
sector to show that the asymptotic dynamics of \tmgl\ can be described
by the conformal symmetry, realized by two independent Virasoro
algebras with different central charges. This result proves that the
hypothesis formulated in \cite{6} is correct, at least at the classical
level.

\section*{Acknowledgements} 

We would like to thank G. Comp\`ere and S. Detournay for useful
remarks. This work was supported by the Serbian Science Foundation
under Grant 141036.

\appendix
\section{Spacelike stretched AdS$_3$} 
\setcounter{equation}{0}

Here, we describe some basic properties of spacelike warped \ads3\
solutions \cite{5,6}.

Maximally symmetric solution of \grl\ for $\L<0$, the anti-de Sitter
space \ads3, is also a solution of \tmgl. It can be represented as a
hypersphere embedded in a four-dimensional flat space $M_4$ with the
metric $\eta=(+,+,-,-)$. By construction, the isometry group
of \ads3\ is $SO(2,2)\sim SL(2,R)\times SL(2,R)$, and we denote the
corresponding Killing vectors by $(\xi_{(0)},\xi_{(1)},\xi_{(2)})$ and
$(\bxi_{(0)},\bxi_{(1)},\bxi_{(2)})$, respectively. After introducing a
convenient set of coordinates $(\t,u,\s)$, analogous to the Euler
angles for the 3-sphere, the metric of \ads3\ can be written in the
form
$$
ds^2=\frac{\ell^2}{4}\left[\cosh\s^2d\t^2-d\s^2
                           -(du+\sinh\s d\t)^2\right]\, ,
$$
where $\{\t,u,\s\}$ are in the range $(-\infty,+\infty)$.

The metric of the spacelike warped $\rm AdS_3$ is given by:
\be
ds^2=\frac{\ell^2}{\n^2+3}\left[\cosh\s^2d\t^2-d\s^2
     -\frac{4\n^2}{\n^2+3}(du+\sinh\s d\t)^2\right]\, ,     \lab{A1}
\ee
where $w:=4\n^2/(\n^2+3)$  is the warp factor. The isometry group
of \eq{A1} is generated by four Killing vectors,
\bea\lab{A2}
&&\xi_{(2)}=2\pd_u\, ,                                     \nn\\
&&\bxi_{(1)}=2\sin\t\tanh\s \pd_\t-2\cos\t\pd_\s
            +\frac{2\sin\t}{\cosh\s}\pd_u\, ,              \nn\\
&&\bxi_{(2)}=-2\cos\t\tanh\s\pd_\t-2\sin\t\pd_\s
            -\frac{2\cos\t}{\cosh\s}\pd_u\, ,              \nn\\
&&\bxi_{(0)}=2\pd_\t\, , \eea which satisfy the commutator algebra
$u(1)\times sl(2,R)$. For $\n^2>1$, we have $w>1$, and the metric
\eq{A1} describes the {\it spacelike stretched\/} \ads3.

One can show that the spacelike stretched \ads3\ is locally
isometric to the black hole \eq{2.4}, by using the following
change of coordinates: \bea &&\t=\arctan\left[\frac{2
\sqrt{(r-r_+)(r-r_-)}\sinh\phi}
                       {2r-r_+-r_-}\right]\, ,             \nn\\
&&\ell u=\frac{\n^2+3}{4\n}\left[2t+\ell\left(\n(r_++r_-)
  -\sqrt{r_+r_-(\n^2+3)}\right)\vphi\right]
  -\ell{\rm atanh}\left[\frac{r_++r_--2r}{r_+-r_-}\coth\phi\right]\nn\\
&&\s={\rm asinh}\left[\frac{2 \sqrt{(r-r_+)(r-r_-)}
     \cosh\phi}{r_+-r_-}\right]\, ,                        \lab{A3}
\eea
where
$$
\phi=\dis\frac{(3+\n^2)(r_+-r_-)}{4\ell}\vphi\, ,
$$
and $(t,r,\vphi)$ are the usual Schwarzschild-like coordinates.
Note that the black hole \eq{2.4} is obtained from the
spacelike stretched \ads3\ by the identification $\vphi\sim\vphi+2\pi$.
Expressed in terms of the new coordinates, the Killing vectors \eq{A2}
take the form:
\bea
\xi_{(2)}&=&\frac{4\n^2}{\n^2+3}\ell\pd_t\, ,              \nn\\
\bxi_{(1)}&=&-2\frac{\sqrt{r_+r_-(\n^2+3)}(r_++r_--2r)
  +2\n\left[r(r_++r_-)-2r_+r_-\right]}
  {(3+\n^2)(r_+-r_-)\sqrt{(r-r_+)(r-r_-)}}\sinh\phi\ell\pd_t\,,\nn\\
  &&-2\sqrt{(r-r_+)(r-r_-)}\cosh\phi\pd_r
    +\frac{4\ell(2r-r_+-r_-)}{(3+\n^2)(r_+-r_-)
      \sqrt{(r-r_+)(r-r_-)}}\sinh\phi\pd_\vphi\, ,         \nn\\
\bxi_{(2)}&=&-4\frac{\sqrt{r_+r_-(\n^2+3)}-\n(r_++r_-)}
  {(3+\n^2)(r_+-r_-)}\ell\pd_t
  -\frac{8\ell}{(3+\n^2)(r_+-r_-)}\pd_\vphi\, ,            \\
\bxi_{(0)}&=&-2\frac{\sqrt{r_+r_-(\n^2+3)}(r_++r_--2r)
                    +2\n\left[r(r_++r_-)-2r_+r_-\right]}
  {(3+\n^2)(r_+-r_-)\sqrt{(r-r_+)(r-r_-)}}\cosh\phi\ell\pd_t\,,\nn\\
&&-2\sqrt{(r-r_+)(r-r_-)}\sinh\phi\pd_r
  -\frac{4\ell(r_++r_--2r)}
        {(3+\n^2)(r_+-r_-)\sqrt{(r-r_+)(r-r_-)}}\cosh\phi\pd_\vphi\,.\nn
\eea
After the identification $\vphi\sim\vphi+2\p$, only $\xi_{(2)}$ and
$\bxi_{(2)}$ remain the Killing vectors of the black hole \eq{2.4}.

The asymptotic form of $\xi_{(2)}$ and $\bxi_{(2)}$ is quite simple,
while for $\bxi_{(1)}$ and $\bxi_{(0)}$ we have:
\bea
\bxi_{(1)}&=&-4\left[\frac{\n(r_++r_-)-\sqrt{(\n^2+3)r_+r_-}}
  {(3+\n^2)(r_+-r_-)}\sinh\phi+\cO_2\right]\ell\pd_t       \nn\\
  &&-(2\cosh\phi r+\cO_0)\pd_r
  +\left[\frac{8\ell}{(3+\n^2)(r_+-r_-)}\sinh\phi
  +\cO_2\right]\pd_\vphi\, ,                               \nn\\
\bxi_{(0)}&=&-4\left[\frac{\n(r_++r_-)-\sqrt{(\n^2+3)r_+r_-}}
  {(3+\n^2)(r_+-r_-)}\cosh\phi+\cO_1\right]\ell\pd_t\, ,   \nn\\
&&-(2r\sinh\phi+\cO_0)\pd_r
  +\left[\frac{8\ell\cosh\phi}{(3+\n^2)(r_+-r_-)}
  +\cO_2\right]\pd_\vphi\, .                               \nn
\eea
These expressions are needed for our discussion of the asymptotic conditions
in section 3.

\section{The curvature, Ricci and Cotton tensors} 
\setcounter{equation}{0}

In this appendix, we present some technical details related to the
form of the spacelike warped black hole solution in the first
order formalism.

Using the connection \eq{2.5b}, we find that the curvature $R_i$ is
given by
\bsubeq
\bea
&&R_0=\left(\b'B+2\b\g\right)b^0b^1
        -\left(\g'B+\g^2+\b^2\right)b^1b^2\, ,             \nn\\
&&R_1= -(\a\g+\b^2)b^2b^0\, ,                              \nn\\
&&R_2=-\left(\a'B+\a^2-3\b^2\right)b^0b^1
        -\left(\b'B+2\b\g\right)b^1b^2\, ,
\eea
or equivalently:
\bea
&&R_0=-\frac{3}{\ell^2}(\n^2-1)NKN_\vphi b^0b^1
      -\frac{1}{\ell^2}\left(\n^2+3(\n^2-1)N^2\right)b^1b^2\, ,\nn\\
&&R_1= -\frac{\n^2}{\ell^2}b^2b^0\, ,                      \nn\\
&&R_2=-\frac{1}{\ell^2}\left(3-2\n^2-3(\n^2-1)N^2\right)b^0b^1
        +\frac 3{\ell^2}(\n^2-1)NKN_\vphi b^1b^2\, ,
\eea
\esubeq
Then, the  components of the Ricci tensor
$\ric_{mn}=-\ve^{kl}{_m}R_{kln}$ are found to be:
\bsubeq
\bea
&&\ric_{00}=\a'B+\a^2+\a\g-2\b^2\,,\qquad \ric_{01}=0\, ,  \nn\\
&&\ric_{11}=-\a'B-\g'B-\a^2-\g^2+2\b^2\,,\qquad \ric_{12}=0\,,\nn\\
&&\ric_{22}=-\g'B-\g^2-\a\g-2\b^2\, ,\qquad \ric_{20}=-(\b'B+2\b\g)\,,
\eea
or equivalently:
\bea
&&\ric_{00}=\frac{1}{\ell^2}\left(3-\n^2-3(\n^2-1)N^2\right)\, ,
\qquad \ric_{01}=0\, ,                                     \nn\\
&&\ric_{11}=-\frac{1}{\ell^2}(3-\n^2)\,,\qquad \ric_{12}=0\,,\nn\\
&&\ric_{22}=-\frac{1}{\ell^2}\left(2\n^2+3(\n^2-1)N^2\right)\,,
\qquad \ric_{20}=\frac{3}{\ell^2}(\n^2-1)NKN_\vphi\, .
\eea
\esubeq
Finally, the Cotton 2-form reads:
\bea
&&C_0=\frac{9\n}{\ell^3}(\n^2-1)NKN_\vphi b^0b^1
      -\frac{3\n}{\ell^3}(\n^2-1)(3N^2-1)b^1b^2\, ,        \nn\\
&&C_1=\frac{3\n}{\ell^3}(\n^2-1) b^2b^0\nn\,,\\
&&C_2=-\frac{3\n}{\ell^3}(\n^2-1)(2+3N^2)b^0b^1
      -\frac{9\n}{\ell^3}(\n^2-1)NKN_\vphi b^1b^2\, .
\eea

\section{Improved asymptotic conditions}
\setcounter{equation}{0}

Our asymptotic conditions \eq{3.2} are chosen so that all higher-order
terms are left completely arbitrary. However, this feature
can be improved by noting that expressions that vanish on shell should
have an arbitrarily fast asymptotic decrease. By applying this
principle to the secondary constraints, we obtain the following
relations between higher-order terms:
\bea
&&\pd_0\Om^0{_2}=\cO_1\,,\quad \pd_0\L^0{_2}=\cO_1\, ,     \nn\\
&&\pd_0 B^2{_2}=\cO_1\,,\quad \pd_0\Om^2{_2}=\cO_1\,,\quad
\pd_0\L^2{_2}=\cO_1\, ,
\eea
and also
\bsubeq\lab{C2}
\bea
&&\frac{\n}{\ell}B^2{_2}+\Om^2{_2}
   +\frac{\sqrt{3(\n^2-1)(\n^2+3)}}{2\ell}
   \left(-\frac{\n}{\ell}B^1{_1}+\Om^1{_1}\right)r^2=\cO_1\, ,    \lab{C2a}\\
&&-\frac{\n}{\sqrt{\n^2+3}}\left(\Om^0{_2}
  +\frac{3}{2a}\L^0{_2}\right)                                    \nn\\
&&+\frac{\sqrt{3(\n^2+3)(\n^2-1)}}{2\ell}
   \left(-\frac{2\n}{\ell}B^1{_1}+\Om^1{_1}\right)r^2=\cO_1\, ,   \lab{C2b}\\
&&\frac{3-2\n^2}{2\ell}B^2{_2}+\n\Om^2{_2}+\frac{3\n}{2a}\L^2{_2} \nn\\
&&+\frac{\sqrt{3(\n^2-1)(\n^2+3)}}{2\ell}
  \left(\frac{3(2\n^2+1)}{2\ell}B^1{_1}
   -\n\Om^1{_1}+\frac{3\n}{2a}\L^1{_1}\right)r^2=\cO_1\,,         \lab{C2c}\\
&&\frac{1}{\ell}B^2{_2}+\frac{(4\n^2+3)}{6\n}\Om^2{_2}
   +\frac{\n}{2a}\L^2{_2}                                         \nn\\
&&+\frac{\sqrt{3(\n^2-1)(\n^2+3)}}{2\ell}\left(\frac{1}{\ell}B^1{_1}
  +\frac{1}{2\n}\Om^1{_1}-\frac{\n}{2a}\L^1{_1}\right)r^2=\cO_1\,,\lab{C2d}\\
&&-\frac{\n}{\sqrt{\n^2+3}}\left(\frac{4\n^2+3}{6\n^2}\Om^0{_2}
  +\frac{1}{2a}\L^0{_2}\right)                                    \nn\\
&&+\frac{\sqrt{3(\n^2+3)(\n^2-1)}}{2\ell}\left(\frac{\Om^1{_1}}3
  +\frac{\L^1{_1}}{2a}\right)r^2=\cO_1\, .                        \lab{C2e}\,.
\eea
From \eq{C2b} and \eq{C2e}, we obtain
\be
-\frac\n{\sqrt{\n^2+3}}\frac{2\n^2+3}{2\n^2}\Om^0{_2}
  +\frac{\sqrt{3(\n^2+3)(\n^2-1)}}{2\ell}\left(\frac{2\n}{\ell}B^1{_1}
  +\frac{3}{2a}\L^1{_1}\right)r^2=\cO_1\, .                     \lab{C2f}
\ee
\esubeq
By eliminating $B^1{_1}$, $\Om^1{_1}$ and $\L^1{_1}$ from
the remaining equations in \eq{C2}, one finds the relation
\be
\frac{\n}{\ell}B^2{_2}+\frac{4}{3}\Om^2{_2}+\frac{1}{a}\L^2{_2}
  +\frac{2\sqrt{\n^2+3}}{3\n}\Om^0{_2}=\cO_1\, ,           \lab{C3}
\ee
that ensures finiteness of the angular momentum.

\section{Hamiltonian and constraints} 
\setcounter{equation}{0}

In this appendix, we present a brief overview of the canonical structure
of \tmgl\ \cite{11}.

Starting with the Lagrangian variables $(b^i{_\m},\om^i{_\m},\l^i{_\m})$ and
the corresponding canonical momenta  $(\pi_i{^\m},\Pi_i{^\m},p_i{^\m})$,
we find the following primary constraints:
\bea
&&\phi_i{^0}:=\pi_i{^0}\approx 0\,,\qquad\,\,
  \phi_i{^\a}:=\pi_i{^\a}-\ve^{0\a\b}\l_{i\b}\approx 0\, , \nn\\
&&\Phi_i{^0}:=\Pi_i{^0}\approx 0\, ,\qquad
  \Phi_i{^\a}:=\Pi_i{^\a}
       -a\ve^{0\a\b}(2b_{i\b}+\m^{-1}\om_{i\b})\approx 0\,.\nn\\
&&p_i{^\m}\approx 0\,.
\eea
The canonical Hamiltonian has the form (up to a 3-divergence):
\bea
&&\cH_c= b^i{}_0\cH_i+\om^i{}_0\cK_i+\l^i{_0}\cT^i\,,       \nn\\
&&\cH_i=-\ve^{0\a\b}\left(aR_{i\a\b}
        -\L\ve_{ijk}b^j{}_\a b^k{}_\b+\nabla_\a\l_{i\b}\right)\,,\nn\\
&&\cK_i=-\ve^{0\a\b}\left(aT_{i\a\b}+a\m^{-1}R_{i\a\b}
        +\ve_{ijk}b^j{}_\a \l^k{}_\b\right) \, ,           \nn\\
&&\cT_i=-\frac{1}{2}\ve^{0\a\b}T_{i\a\b}\,.                \nn
\eea

After constructing the total Hamiltonian $\cH_T$, the
consistency requirements on the primary constraints produce
the secondary constraints,
\be
\cH_i\approx 0\, ,\qquad \cK_i\approx 0\, ,\qquad \cT_i\approx 0\, ,
\ee
and yield the additional relations which determine the multipliers
$u^i{_\a},v^i{_\a}$ and $w^i{_\a}$. The modified total Hamiltonian
takes the form (up to a 3-divergence):
\bea
&&\cH'_T=b^i{}_0\bcH_i+\om^i{}_0\bcK_i+\l^i{_0}\bcT_i
         +u^i{}_0\pi_i{}^0+v^i{}_0\Pi_i{}^0+w^i{_0}p_i{^0}\, ,\nn\\
&&\bcH_i=\cH_i-\nabla_\b\phi_i{}^\b
  -\frac{\m}{2a}\ve_{ijk}\l^j{_\b}\Phi^{k\b}
  +\ve_{ijk}\left(2\L b^j{_\b}+\m\l^j{_\b}\right)p^{k\b}\,,\nn\\
&&\bcK_i=\cK_i-\ve_{ijk}b^j{}_\b\phi^{k\b}-\nabla_\b\Phi_i{}^\b
  -\ve_{ijk}\l^j{_\b}p^{k\b}\, ,                           \nn\\
&&\bcT_i=\cT_i
  -\frac{\m}{2a}\ve_{ijk} b^j{_\b}\Phi^{k\b}-\nabla_\b p_i{^\b}
  +\m\ve_{ijk}b^j{}_\b p^{k\b}\, .                         \nn
\eea

The consistency conditions of the secondary constraints lead to three
independent tertiary constraints,
\bsubeq
\bea
&&\th_{0\b}:=\l_{0\b}-\l_{\b 0}\approx 0\, ,               \\
&&\th_{\a\b}:=\l_{\a\b}-\l_{\b\a}\approx 0\, ,
\eea
\esubeq
while the consistency of $\th_{\a\b}$ yields a new, quartic constraint:
\be
\Psi=3\L+\m\l\approx 0\, .
\ee
Further consistency requirements determine the multipliers
$w^i{_0}{}':=w^i{_0}-u^k{_0}\l_k{^i}$, whereby the consistency
procedure is completed. The final form of the total Hamiltonian
reads:
\bea
\hcH_T&=&\bar\cH_T
  +u^i{_0}\pi_i{^0}{}'+v^i{_0}\Pi_i{^0}\, ,                \\
\bar\cH_T&:=&b^i{_0}\bcH_i+\om^i{_0}\bcK_i+\l^i{_0}\bcT_i
             +\bw_{\b 0}'p^{\b 0}+\bw_{00}'p^{00}\, .      \nn
\eea
where $\pi_i{^0}{}':=\pi_i{^0}+\l_i{^k}p_k{^0}$, and the multipliers
with an overbar are determined.

As far as the classification of constraints is concerned, we find that
$\pi_i{^0}{}',\Pi_i{^0}$ and
\bea
&&\hcH_i=\bcH_i+\l_i{^k}\bcT_k
          +h_i{^\r}(\nab_\r \l_{jk})b^k{_0}p^{j0}\, ,      \nn\\
&&\hcK_i=\bcK_i-\ve_{ijk}(\l^j{_0}p^{k0}-b^j{_0}\l^k{_n}p^{n0})\,,
\eea
are first class, while all the others are second class.

The canonical generator of gauge transformations has the form \eq{4.1}.


\begin{thebibliography}{99} 

\bibitem{1} S. Deser, R. Jackiw and S. Templeton,
  Three-Dimensional Massive Gauge Theories, Phys. Rev. Lett.
  {\bf 48} (1982) 975; Topologically Massive Gauge Theories,
  Ann. Phys. {\bf 140} (1982) 372.\vsm

\bibitem{2} M. Ba\~nados, C. Teitelboim and J. Zanelli, The Black Hole
  in Three-Dimensional Spacetime, Phys. Rev. Lett. {\bf 16} (1993)
  1849;\\
  M. Ba\~nados, M. Henneaux, C. Teitelboim and J. Zanelli, Geometry of
  2+1 Black Hole, Phys. Rev. D {\bf 48} (1993) 1506 .\vsm

\bibitem{x1} K. Ait Moussa, G. Clement and C. Leygnac, The black
holes in topologically massive gravity, Class. Quant. Grav. {\bf 20}
(2003) L277.\vsm

\bibitem{3} W. Li, W. Song and A. Strominger, Chiral Gravity in
  Three Dimensions, JHEP {\bf 0804} (2008) 082.\vsm

\bibitem{4} A. Strominger, A Simple Proof of the Chiral Gravity Conjecture,\\
  preprint arXiv:0808.0506v1[hep-th];\\
  A. Maloney, W. Song and A. Strominger, Chiral Gravity,
  Log Gravity and Extremal CFT, arXiv:0903.4573 [hep-th].\vsm

\bibitem{5} I. Bengtsson and P. Sandin,
  Anti de Sitter Space, squashed and stretched, preprint
  arXiv:gr-qc/0509076.\vsm

\bibitem{6} D. Anninos, W. Li, M. Padi, W. Song and A. Strominger,
  Warped AdS$_3$ black holes, arXiv:0807.3040[hep-th].\vsm

\bibitem{7} D. Anninos, M. Esole and M. Guica, Stability of warped \ads3\
  vacua of topologically massive gravity, arXiv:0905.2612 [hep-th].\vsm

\bibitem{8} D. Anninos, Sailing from Warped \ads3 to Warped dS$_3$ in
  Topologically Massive Gravity, arXiv:0906.1819 [hep-th].\vsm

\bibitem{9} G. Comp\`ere and S. Detournay, Semi-classical central
  charge in topologically massive gravity, Class. Quant. Grav.
  {\bf 26} (2009) 012001.\vsm

\bibitem{10} G. Comp\`ere and S. Detournay,
  Boundary conditions for spacelike and timelike warped \ads3\
  spaces in topologically massive gravity, arXiv:0906.1243.\vsm

\bibitem{11}  M. Blagojevi\'c and B. Cvetkovi\'c, Canonical structure of
  topologically massive gravity with a cosmological constant,
  JHEP {\bf 0905} (2009) 073;\vsm

\bibitem{12}  J. D. Brown and M. Henneaux, Central Charges in the
  Canonical Realization of Asymptotic Symmetries: An Example from Three
  Dimensional Gravity, Comm. Math. Phys. {\bf 104} (1986) 207;\\
  M. Blagojevi\'c and B. Cvetkovi\'c, Canonical structure of
  3D gravity with torsion, in: {\it Progress in General Relativity and
  Quantum Cosmology\/}, vol. 2, ed. Ch. Benton (Nova Science Publishers,
  New York, 2006), p. 103 (preprint gr-qc/0412134).\vsm

\bibitem{13} H. Sugawara, A field theory of currents,
  Phys. Rev. {\bf 170} (1968) 1659.\vsm

\bibitem{14} M. Blagojevi\'c, {\it Gravitation and gauge symmetries\/}
  (IoP Publishing, Bristol, 2002);\\
  T. Ort\'in, {\it Gravity and strings\/}, (Cambridge University Press,
  Cambridge, 2004).\vsm

\bibitem{x2} A. Bouchareb and G. Cl\'ement, Black hole mass and
  angular momentum in topologically massive gravity, Class. Quant.
  Grav. {\bf 24} (2007) 5581-5594.\vsm

\bibitem{15} S. N. Solodukhin, Holography with gravitational
  Chern-Simons, Phys. Rev. D{\bf 74} (2006) 024015;\\
  Y. Tachikawa, Black hole entropy in the presence of Chern-Simons
  terms, Class. Quant. Grav. {\bf 24} (2007) 737-744.\vsm

\bibitem{16} M. Henneaux and C. Teitelboim, Asymptotically anti-de Sitter
  spaces, Comm. Math. Phys. {\bf 98} (1985) 391.\vsm

\bibitem{17} L. Castellani, Symmetries of constrained Hamiltonian
  systems, Ann. Phys. (N.Y.) {\bf 143} (1982) 357.\vsm

\bibitem{18} T. Regge and C. Teitelboim, Role of surface integrals in the
  Hamiltonian formulation of general relativity,
  Annals Phys. {\bf 88} (1974) 286.\vsm

\bibitem{19} S. Deser and B. Tekin, Energy in topologically massive
gravity, Class. Quantum Grav. {\bf 20} (2003) L259-L262.\vsm

\bibitem{20} J. D. Brown and M. Henneaux, On the Poisson bracket of
  differentiable gene\-ra\-tors in classical field theory,
  J. Math. Phys. {\bf 27} (1986) 489.\vsm

\end{thebibliography}
\end{document}